# Stored Luminescence Computed Tomography

Wenxiang Cong, Chao Wang, and Ge Wang*

Biomedical Imaging Center/Cluster

Center for Biotechnology and Interdisciplinary Studies

Department of Biomedical Engineering

Rensselaer Polytechnic Institute, Troy, NY 12180, USA

congw@rpi.edu, chaowang9@yahoo.com, ge-wang@ieee.org

**Abstract**

The phosphor nanoparticles made of doped semiconductors, pre-excited by well-collimated X-ray radiation, were recently reported for their light emission in the range 650–770 nm upon NIR light stimulation. The characteristic of X-ray energy storage and NIR stimulated emission is highly desirable to design targeting probes and improve molecular and cellular imaging. Here we propose "stored luminescence computed tomography" (SLCT), perform realistic numerical simulation, and demonstrate a much-improved spatial resolution in a preclinical research context. The future opportunities are also discussed along this direction.

## 1. Introduction

X-ray luminescence CT (XLCT) is a hot topic for preclinical imaging [1-3], which combines X-ray and optical imaging to improve image resolution relative to purely optical imaging modalities such as fluorescence tomography [4, 5] and bioluminescence tomography [6, 7]. XLCT utilizes X-ray luminescent nanophosphors (NPs) as imaging probes. NPs can be excited with a pencil, fan or cone beam of X-rays, and the NP luminescence can be readily generated and efficiently collected using a sensitive light detection system. XLCT is an analog to fluorescence diffuse optical tomography (FDOT) but the former has an advantage over the latter in terms of penetration depth and photo-stability. However, this type of luminescent light signals comes from all excited NPs and is highly diffusive, limiting the accuracy of image reconstruction.

Recently, $MgGa_2O_4:Cr^{3+}$ nanophosphors, as well as others, were reported to be capable of restraining luminescence emission for ten hours or longer after one-time X-ray excitation [8, 9]. The stored energy is released upon NIR light stimulation at subsequent time points leads to *effectively controlled* luminescence emission in the range of 650–770 nm. The mechanism for the stored luminescence in such nanophosphors lies in the modified energy diagram of the doped semiconductor

MgGa$_2$O$_4$:Cr$^{3+}$ of nearly 44% cationic site inversion due to nominal Mg deficiency. The dopant Cr$^{3+}$ ions occupy octahedral sites which are associated with the stored luminescence emission spectrum around 707 nm (corresponding to the dopant associated $^2$E($^2$G) → $^4$A$_2$($^4$F) transition); please see Figure A-1 in the appendix for more details. The advent of such nanophosphors opens a door to significant protocol flexibility and performance gain for molecular and cellular imaging.

Here we propose "stored luminescence computed tomography" (SLCT), perform realistic numerical simulation, and demonstrate a much-improved spatial resolution in a preclinical research context. SLCT is an analog to XLCT, but the former outperforms the latter because stored luminescence can be readout in a much more sophisticated way than instantaneous emitted luminescence on which XLCT relies. As shown in this paper, the freedom for selective data acquisition implies a major improvement in image resolution, holding a great potential for sensitive and specific small animal imaging in general.

## 2. Imaging Principles

MgGa$_2$O$_4$:Cr$^{3+}$ nanoparticles can be functionalized to target specific cells and then introduced into an object such as a living mouse [8, 10]. X-rays from an x-ray tube are collimated into a narrow beam such as a pencil or fan beam for excitation of regions of interest (ROI) in the mouse. Part of the x-ray energy is then deposited in the nanophosphors. When stimulated by NIR laser light in various patterns, the pre-excited nanophosphors will emit luminescence photons in the range 650–770 nm. The retrospectively stimulated luminescence emission data are collected on the surface of the mouse for tomographic image reconstruction.

SLCT imaging is to localize and quantify a distribution of energy-storing nanophosphors such as MgGa$_2$O$_4$:Cr$^{3+}$ nanoparticles in a 3D object such as a mouse. Since the measurement of NIR light signals can be spatially and temporally resolved by stimulating energy-storing nanophosphors at any location on the surface of the object and any time instant as long as the energy is kept in the nanoparticles, the resultant dataset will carry more tomographic information than the counterpart in a corresponding XLCT experiment. Generally speaking, many problems of optical tomography are underdetermined with possible false solutions due to the inherent non-uniqueness and data noise [7, 11]. Interestingly, this ill-posedness can be effectively overcome with informative combinations of well-defined x-ray beam shapes and NIR stimulation patterns.

It is underlined that two powerful imaging features are unique to SLCT, which are impossible with other optical imaging modalities such as XLCT. First, a field of view (FOV) for a SLCT study can be clearly defined with X-rays. This means not only a direct energy distribution from a single X-ray beam but also a synthetic energy distribution from multiple X-ray beams. The latter scheme is an analog to tomotherapy for radiation oncology. Second, an energy distribution carried by energy storing nanoparticles can be optically read out in a multiplexing fashion. NIR light patterns can be projected externally anywhere around the FOV in a time sequence, which is more effective than measuring all luminescent signals simultaneously from the distribution as a whole. The NIR stimulation patterns can be coded in different frequencies as well. The light penetration depth depends on its spectrum and intensity. For example, SLCT imaging can be performed in an onion peeling fashion. In other words, external shells can be initially stimulated and reconstructed, depleting X-ray energies in these shells. Then, the subsequent signals must come from internal shells. With such an onion peeling strategy, the FOV can be step-by-step shrunk for more accurate and more reliable image reconstruction, which is impossible with XLCT, fluorescence tomography and bioluminescence tomography.

## 2.1. X-ray Excitation

Incident X-rays can be easily collimated into a narrow beam to excite energy-storing nanoparticles such as $MgGa_2O_4:Cr^{3+}$ in a living mouse. The X-ray intensity distribution $I(\mathbf{r})$ in the animal can be described by the Lambert-Beer law:

$$X(\mathbf{r}) = X_0 \exp\left(-\int_0^{|\mathbf{r}-\mathbf{r_0}|} \mu\left(\mathbf{r_0} + t \cdot \frac{\mathbf{r}-\mathbf{r_0}}{|\mathbf{r}-\mathbf{r_0}|}\right) dt\right), \tag{1}$$

where $\mathbf{r_0}$ is a source position, $X_0$ the incident x-ray intensity, and $\mu$ the linear attenuation coefficient [mm$^{-1}$] which can be computed with X-ray computed tomography (CT). Energy stored in the nanoparticles is determined by local X-ray flux intensities. Therefore, we need to compute a stored X-ray energy density distribution inside the animal according to Eq. (1). As mentioned above, a more desirable X-ray energy distribution can be synthesized using a tomotherapy approach [12]. The literature on tomotherapy is extensive. An established radiation therapeutic planning technique, such as [13], can be directly applied to deposit a pre-specified X-ray energy distribution such as targeting an ROI.

## 2.2. NIR Stimulation

NIR light is moderately absorbed and strongly scattered in biological tissues. The diffusion approximation (DA) model is commonly used to describe the NIR light propagation in this scenario [6, 14]:

$$-\nabla \cdot [D(\mathbf{r})\nabla\Phi(\mathbf{r})] + \mu_a(\mathbf{r})\Phi(\mathbf{r}) = S(\mathbf{r}), \quad \mathbf{r} \in \Omega, \qquad (3)$$

where $\mathbf{r}$ is a position vector, $\Phi(\mathbf{r})$ an NIR fluence rate [Watts/mm$^2$], $S(\mathbf{r})$ an NIR source [Watts/mm$^3$], $\mu_a$ the absorption coefficient [mm$^{-1}$], $D$ the diffusion coefficient defined by $D = [3(\mu_a + \mu'_s)]^{-1}$, $\mu'_s$ the reduced scattering coefficient [mm$^{-1}$], and $\Omega \subset R^3$. If no NIR photon travels across the boundary $\partial\Omega$ into the tissue domain $\Omega$, the DA is constrained by the Robin boundary condition

$$\Phi(\mathbf{r}) + 2\alpha D(\mathbf{r})(\nu \cdot \nabla\Phi(\mathbf{r})) = 0, \quad \mathbf{r} \in \partial\Omega, \qquad (4)$$

where $\nu$ is the outward unit normal vector on $\partial\Omega$, and $\alpha$ the boundary mismatch factor. The boundary mismatch factor between the tissue of a refractive index $n$ and the air can be approximated by $\alpha = (1+\gamma)/(1-\gamma)$ with $\gamma = -1.4399n^{-2} + 0.7099n^{-1} + 0.6681 + 0.0636n$ [15]. Thus, the measurable exiting photon flux on the surface of the animal is expressed as

$$m(\mathbf{r}) = -D(\mathbf{r})(\mathbf{v} \cdot \nabla\Phi(\mathbf{r})), \quad \mathbf{r} \in \partial\Omega. \qquad (5)$$

The intensity of the NIR luminescence emission depends on the density of energy-storing nanoparticles $\rho(\mathbf{r})$, the X-ray intensity $X(\mathbf{r})$, the laser intensity $L(\mathbf{r})$, and the stored luminescence photon yield $\varepsilon$ of the nanoparticles, which can be defined as the quantum yield per a unit nanoparticle density. Although this dependency is nonlinear, it is assumed in this feasibility study that the intensity of the stimulated light emission, $S(\mathbf{r})$, is linearly proportional to each of the involved densities:

$$S(\mathbf{r}) = \varepsilon X(\mathbf{r})L(\mathbf{r})\rho(\mathbf{r}), \qquad (6)$$

where the stimulating laser intensity distribution in the animal can be calculated by Eq. (3) with a known laser source projected on the surface of the animal. When all the involved quantities are small, the linear system model should work well.

### 2.3. Discretization

Eqs. (3)-(4) can be discretized into a matrix equation linking the nanoparticle distribution $\rho$ and the NIR photon fluence rate $\Phi(\mathbf{r})$ at a node $\mathbf{r}$ using the finite element analysis [14]:

$$\mathbf{A} \cdot \Phi = \mathbf{F} \cdot \rho, \tag{7}$$

where the elements of the matrix $\mathbf{A}$ are

$$a_{ij} = \int_\Omega D(\mathbf{r}) \nabla \varphi_i(\mathbf{r}) \cdot \nabla \varphi_j(\mathbf{r}) d\mathbf{r} + \int_\Omega \mu_a(\mathbf{r}) \varphi_i(\mathbf{r}) \varphi_j(\mathbf{r}) d\mathbf{r} + \int_{\partial \Omega} \varphi_i(\mathbf{r}) \varphi_j(\mathbf{r}) / 2\alpha \, d\mathbf{r}, \tag{8}$$

and the elements of the matrix $\mathbf{F}$ are

$$f_{ij} = \varepsilon \int_\Omega X(\mathbf{r}) L(\mathbf{r}) \varphi_i(\mathbf{r}) \varphi_j(\mathbf{r}) d\mathbf{r}. \tag{9}$$

where $\varphi_i$ $(i = 1, 2, , \cdots)$ are the element shape functions. Since the matrix $\mathbf{A}$ in Eq. (7) is positive definite, we have

$$\Phi = (\mathbf{A}^{-1} \mathbf{F}) \cdot \rho. \tag{10}$$

## 2.4. Image Reconstruction

In the compressive sensing (CS) framework, an image can be reconstructed from far less samples than what the Nyquist sampling theorem requires [16]. Based on the image characteristics encountered in the biomedical imaging applications, targeting nanoparticles are often attached to cells of a preferred type and accumulated locally, forming a sparse and/or smooth distribution of nanoparticles. Using a CS technique [17], we can reconstruct a nanophosphor concentration distribution by solving the following optimization problem

$$\begin{aligned} minimum \quad & \|\rho\|_1 \\ subject\ to \quad & (\mathbf{A}^{-1} \mathbf{F}) \cdot \rho = \Phi \\ & \rho \geq 0 \end{aligned} \tag{5}$$

The $l_1$ term is to induce the solution sparsity. An interior-point method can be applied to solve a large-scale $l_1$-regularized optimization problem Eq. (5), aided by the preconditioned conjugate gradient direction. An important property of the $l_1$-regularized optimization is that a bound can be computed on the sub-optimality of $\rho$.

## 3. Numerical simulation

## 3.1. Imaging System

The simulated imaging system consists of a high sensitive EMCCD camera, a mirror imaging device, an X-ray tube, an X-ray collimator, a laser diode, and a sample stage. The EMCCD camera (iXon3 897, Andor Technology) has a 512×512 resolution, 16μm×16μm pixel size, and >90% QE. As shown in Fig. 1, the camera faces the two mirrors, and the sample stage mounted on a motorized linear stage for focal plane adjustment. The two-mirror position is adjusted by another linear stage. A mouse with a lie prostrate position is caged in a transparent box positioned between the two mirrors, which is easy to implement for the in vivo experiments. The two-mirror configuration is incorporated to expand the field of view of the camera and acquire two views of the mouse simultaneously. The entire imaging system is housed within an optically opaque box.

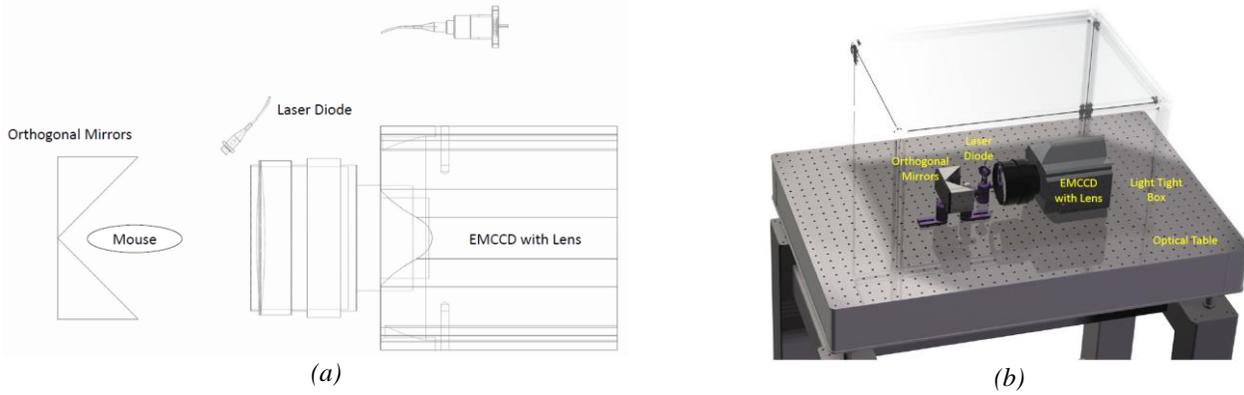

*(a)* *(b)*

***Fig. 1.*** *Rendering of a stored luminescence CT (SLCT) system for small animal imaging. (a) The optical components, (b) an overview of the system (Dr. Fenglin Liu in our group made the system illustration).*

### 3.2. Numerical Simulation

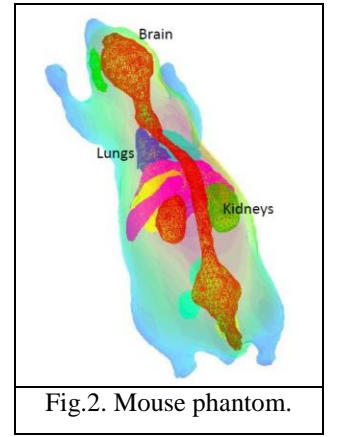

Fig.2. Mouse phantom.

We performed representative numerical tests to evaluate the proposed SLCT approach with a digital mouse phantom [cite ref for the phantom]. As shown in Fig. 2, the mouse phantom was established from the CT slices of a mouse using Amira (Amira 4.0, Mercury Computer Systems, Inc. Chelmsford, MA, USA). The phantom was discretized into 306,773 tetrahedral elements with 58,244 nodes. The stored nanophosphor luminescence re-emission peak was chosen as 700 nm, which was based on the emission characteristics of $MgGa_2O_4 : Cr^{3+}$ [8, 9]. The luminescence photon yield $\varepsilon$ was assumed to be 0.15cm$^3$/mg for 50keV [1]. Appropriate optical parameters were assigned to the mouse model accordingly. The reduced scattering coefficient $\mu'_s(\lambda)$ relies on the stored luminescence emission wavelength $\lambda\,(nm)$ and is approximated by an empirical function:

$$\mu'_s(\lambda) = 10a \cdot \lambda^{-b} \qquad (14)$$

where a and b are the constants depending on the tissue type. The organ-specific values for a and b can be found in [18]. The tissue absorption depends on the local oxy-hemoglobin (HbO$_2$), deoxy-hemoglobin (Hb) and water (W) concentrations. The absorption coefficient $\mu_a(\lambda)$ can be approximated as the weighted sum of the three absorption coefficients $\mu_{aHbO_2}(\lambda)$, $\mu_{aHb}(\lambda)$ and $\mu_{aW}(\lambda)$, which were calculated from the corresponding absorbance spectra reported in [18]:

$$\mu_a(\lambda) = S_B[x\mu_{aHb}(\lambda) + (1-x)\mu_{aHbO_2}(\lambda)] + S_W \mu_{aW}(\lambda) \tag{15}$$

where $x = HbO_2/(HbO_2 + Hb)$ is the ratio between oxy-hemoglobin and total hemoglobin concentration, $S_B$ and $S_W$ are scaling factors. The nanophosphor concentration distribution was set in the mouse lung region from 5µg/mL to 25µg/mL. Figures 3 (a) or 4 (a) show the nanophosphor clusters at the cross section of z=10mm in the phantom.

**X-ray luminescence computed tomography (XLCT):** X-ray luminescence CT (XLCT) is a synergistic imaging modality defining a permissible source region with X-rays to help reconstruct a nanophosphors (NPs) distribution. These conventional NPs are instantaneously excited by common medical X--rays, and the luminescent data are efficiently collected using a highly sensitive light camera. In the first generation CT scanning mode, XCLT uses an x-ray pencil beam excitation, the emitted light can be measured as a line integral. The classic filtered backprojection method can be used to reconstruct an image, with image resolution being decided by the x-ray pencil beam aperture [1, 2]. This scanning mode needs long data acquisition time and is not practical for most preclinical applications. To shorten the scanning time, a cone beam x-ray luminescence computed tomography strategy was proposed [19]. In the cone beam stimulation mode, the X-rays illuminate the whole sample to stimulate all the nanophosphors, and a CCD camera acquires luminescent photons for tomographic imaging. Clearly, this cone beam scanning mode does not sufficiently utilize the primary benefit of XLCT in terms of a reduced permissible source region.

Here, we present a fan-beam stimulation mode for XLCT, which uses a fan-beam of X-rays to irradiate an object such as a mouse, and the nanoparticles on a cross-section of the mouse emit NIR light. The measured NIR light signal (2D) on the external surface of the object is used to reconstruct a nanoparticle distribution (2D) on the excited cross-section. Theoretically, the dimensionality of measure information matches that of the unknown image. The fan-beam scanning mode is the optimal balance between the pencil beam mode and the cone-beam mode for XLCT in terms of imaging efficiency and image quality, and will be focused on in this project to show the advantages of SLCT.

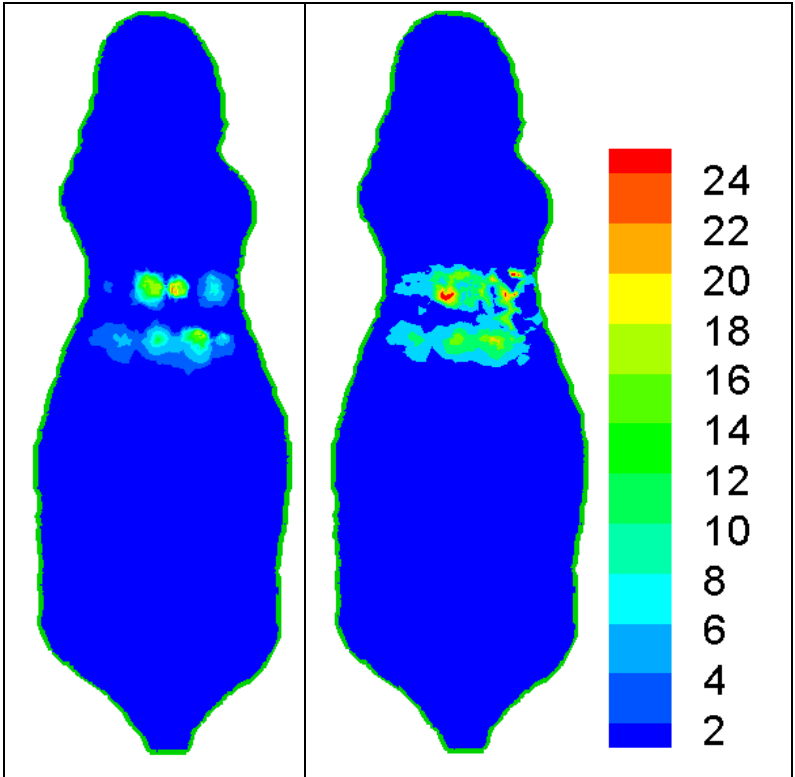

Fig. 3. Comparison between the true nanophosphor distribution and the XLCT reconstruction (Unit: pico Watts/mm$^3$) (a) The true nanophosphor concentration distribution in the mouse phantom; and (b) the reconstructed nanophosphor concentration distribution using XLCT in the fan-beam mode.

An X-ray tube was operated at 50 KeV and 30 mA and collimated into a fan beam with a 1mm thickness. A cross-section was excited at the transverse position of 10mm of the mouse phantom described above. The NIR light emission from the excited nanoparticles was recorded on the surface of the phantom by the CCD camera. The collected NIR data were then corrupted by 5% Gaussian noise to simulate practical conditions. A reconstruction method comparable to what has been proposed in Section 2 was employed to reconstruct the nanophosphor distribution from the NIR data. The reconstructed image revealed the accumulation of the nanophosphors, and generated a 37% relative error, which was defined as $\frac{1}{Num(\rho_k^t > bg)} \sum_{\rho_k^t > bg} \frac{|\rho_k^r - \rho_k^t|}{\max_k(\rho_k^t)}$, where $\rho_k^t$ and $\rho_k^r$ are the true and reconstructed densities on the k-th mesh element respectively, $bg$ was assigned as the background value, and $Num(\rho_k^t > bg)$ the number of elements in the set $\{k : |\rho_k^t > bg\}$. Fig. 3 presents a comparison between the true and reconstructed nanophosphor distributions.

**Stored Luminescence Computed Tomography (SLCT):** The same mouse phantom and parameters settings were used to evaluate the performance of the proposed SLCT approach. The SLCT experiment was performed in an onion peeling fashion. The phantom was stimulated with laser radiation shell by shell. That is, peripheral regions were first stimulated for "cleaning-up", and the data were collected for recursive image reconstruction. The NIR emission signals were gradually collected synchronized to stepwise NIR light stimulation (but still within a very short time window), instead of being collected once for all as for XLCT. This divide-and-conquer data acquisition procedure ensures that the number of unknowns can be significantly reduced to improve spatial resolution and stability of the image reconstruction. To implement this procedure, two laser beams of 650nm stimulate the phantom. One laser beam has the same illumination direction as that of the stimulating X-ray beam, while the other laser beam came from the other side of the phantom in the opposite direction of the first laser beam to stimulate the phantom simultaneously. At the first step, the NIR laser light stimulation mainly excited the peripheral region of the phantom to read-out the nanophosphors near the surface of the phantom with a highly sensitive CCD camera. When the NIR signal acquisition reached a sufficient signal-to-noise ratio, the camera was recorded the first set of NIR light signals for the first shell tomographic imaging. Then, the second shell was similarly approached, so on and so forth. The simulated NIR data on the phantom surface were also corrupted by 5% Gaussian noise. The reconstruction method proposed in Section 2 was step-wise/shell-wise employed to reconstruct the nanophosphor distribution from the NIR data. The algorithm gave an excellent performance in terms of convergence and stability. The reconstructed images are in a close agreement with the truth, as shown in Fig. 4., and the averaged relative error of the reconstructed nanoparticle distribution was less than 22%.

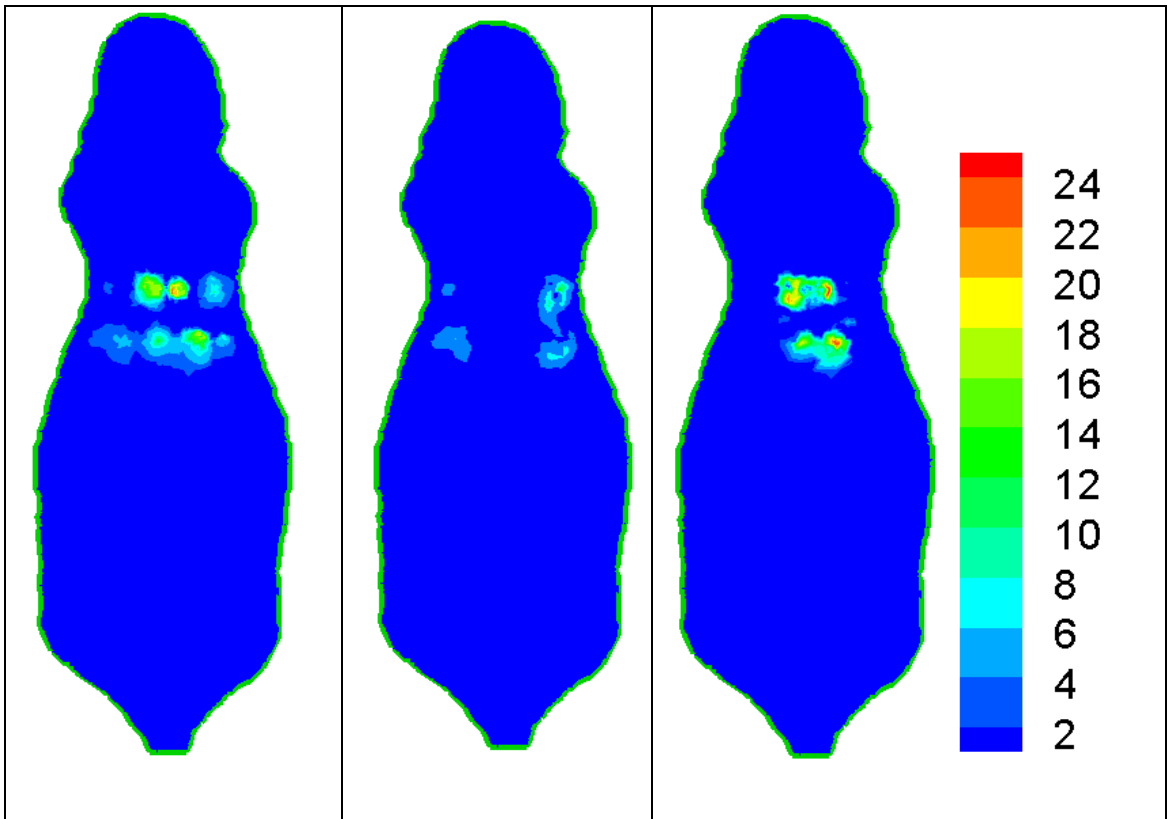

Fig.4. Comparison between the SLCT reconstruction and true nanophosphor concentration distributions (Unit: pico Watts/mm$^3$: (a) True nanophosphor concentration distribution in the numerical mouse phantom; (b) the reconstructed nanophosphor concentration distributions in the peripheral region in the object at the first stage of NIR stimulation; (c) the reconstructed nanophosphor concentration distributions in the central region in the object at the second stage of NIR stimulation.

It can be observed in Figs. 3 and 4 that SLCT is more accurate and more stable than XLCT, as shown in Table 1.

Table 1: Performance comparison between SLCT and XLCT (SLCT performance can be further improved with an optimized NIR excitation scheme).

| Modality | X-ray Stimulation | NIR Excitation | Probe | Stability | Accuracy Error |
|---|---|---|---|---|---|
| **SLCT** | Fan beam | Applied | Nanophosphors | Strong | 22% |
| **XLCT** | Fan beam | No | Nanophosphors | Week | 37% |

## 4. Discussions and Conclusion

We have proposed the stored luminescence tomographic imaging modality which is based on image reconstruction from the shrunk ROIs that physically results from stepwise peripheral clean-ups through stored luminescence re-emissions upon NIR laser stimulations. Owing to deep penetration of well-collimated X-ray excitation as well as NIR laser stimulations, the peripheral clean-ups and shrunk ROIs can be thoroughly and flexibly achieved, making the imaging methodology applicable to high resolution image reconstruction. It is advisable to consider various X-ray excitation patterns and NIR light stimulation approaches for specific purposes in the image reconstruction. For convenience of raw data acquisition and image reconstruction, this modality can achieve stepwise shrunk regions of interest (ROIs) in a 3D object through one-time X-ray excitation followed by successive NIR light stimulations. This stepwise treatment could satisfy accurate and reliable image reconstruction of nanophosphor distribution. Numerical simulations demonstrated the feasibility of the proposed approaches. The stored luminescence tomographic imaging modality may find its pre-clinical applications in monitoring drug delivery and assessing cancer therapy.

Appendix: Energy diagram of doped semiconductors for stored nanophosphor luminescence.

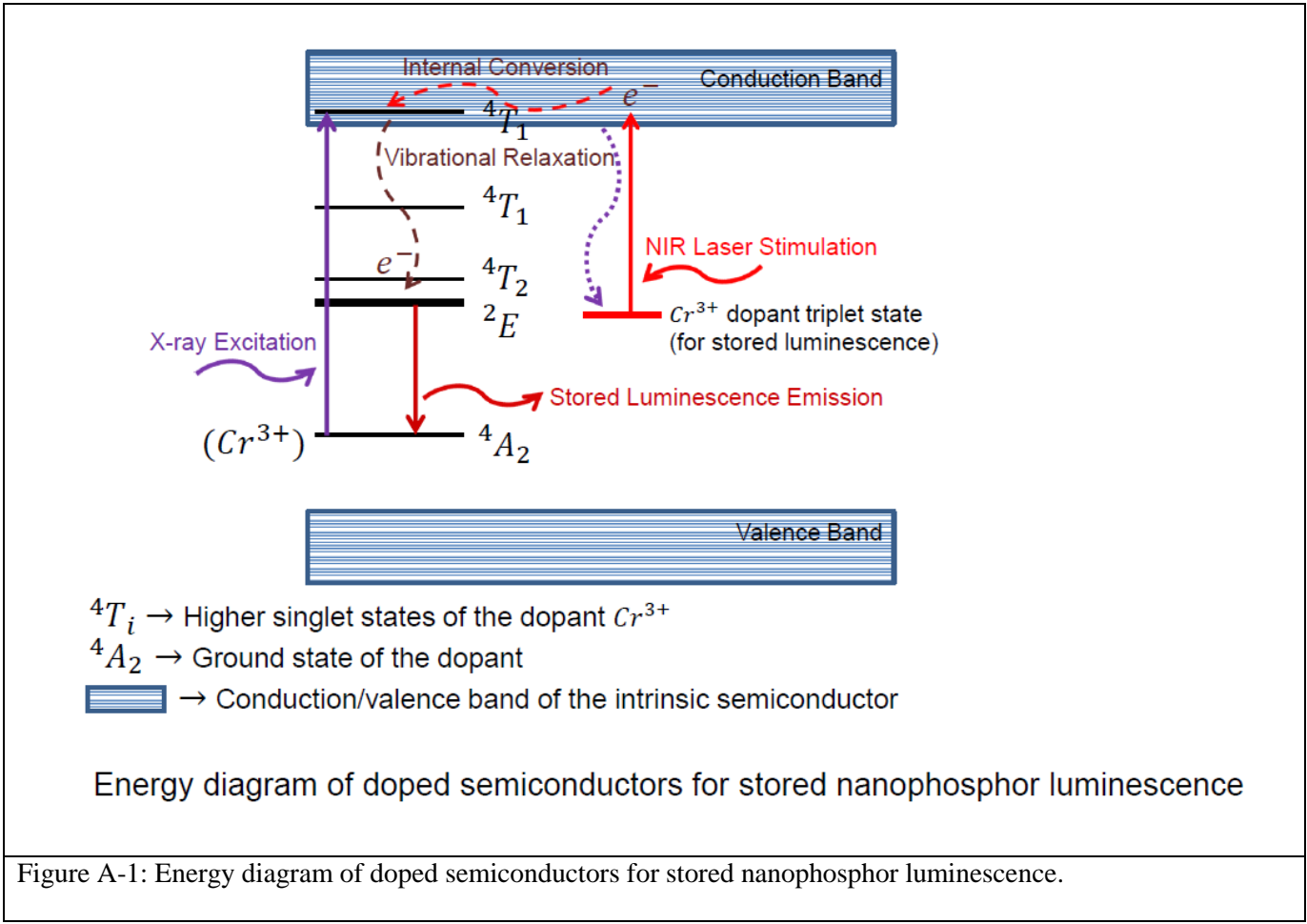

Figure A-1: Energy diagram of doped semiconductors for stored nanophosphor luminescence.